\documentstyle[floats,epsfig,aps,prb]{revtex}

\newcommand{\dfrac}[2]{\frac{\strut \displaystyle{#1}}{\displaystyle{#2}}}

\begin{document}

\draft

\twocolumn[\hsize\textwidth\columnwidth\hsize\csname 
@twocolumnfalse\endcsname

\title{\large\bf Ferromagnetism in the Hubbard model with orbital degeneracy 
in infinite dimensions}

\author{Tsutomu Momoi and Kenn Kubo}
\address{Institute of Physics, University of Tsukuba, 
Tsukuba, Ibaraki 305-8571, Japan}
\date{\hspace*{5cm}}

\maketitle

\begin{abstract}
We study the ferromagnetism due to orbital degeneracy in the 
Hubbard model in infinite dimensions. 
The model contains the intraorbital repulsion $U$, 
the interorbital repulsion $U^\prime$, the exchange $J$ 
(Hund coupling) and the pair hopping $J^\prime$, where 
all of them originate from the on-site Coulomb interaction. 
The ground state of the effective one-site problem was obtained 
by exact diagonalizations.
At the $1/4$-filling, we found two insulating phases; one is a 
ferromagnetic phase with alternating orbital order and the other 
is antiferromagnetic with uniform orbital order. If
 electrons are doped into the $1/4$-filling, the ferromagnetic phase
 still survives and becomes metallic, while the antiferromagnetic phase
disappears. 
This result indicates that the double-exchange mechanism 
is relevant to stabilize metallic ferromagnetism  in
infinite dimensions. 
\end{abstract}
\vskip0.3pc]
\narrowtext

Many investigations have been done on the Hubbard model, to clarify 
whether the Coulomb interaction can realize ferromagnetism
in itinerant electron systems. 
They revealed that the simple single-band 
Hubbard model on the hypercubic lattice does not easily
show ferromagnetism and that some additional properties are necessary 
to stabilize ferromagnetism. 
There are several proposals for such properties. One is the special 
lattice structure (or band structure) which favors ferromagnetism. 
For example, ferromagnetism on
the fcc lattice\cite{Kanamori} and of the flat (or nearly flat)-band 
model\cite{Mielke,Tasaki} are discussed in this context. 
Another is the existence of the orbital degeneracy.\cite{Slater,Zener,VVleck} 
It was argued that the on-site Hund coupling between electrons in degenerate 
orbitals creates indirect ferromagnetic coupling between conduction 
electrons.\cite{VVleck} 
 
Numerical study of strongly correlated electron systems 
is quite difficult in higher dimensions than one.
In the infinite-dimensional ($d=\infty$) limit, however, it is manageable with 
the aid of recently developed technique.\cite{GeorgesKKR} 
In this limit we can treat quantum fluctuations completely, by taking 
local interactions into account exactly, 
and we can neglect spatial fluctuations. 
Monte Carlo simulations of the $d=\infty$ single-band Hubbard
model revealed that any ferromagnetic phase does not appear on 
the hypercubic lattice\cite{FreericksJ} and, on the other hand,
metallic ferromagnetism occurs on the fcc-type lattice.\cite{Ulmke} 
These two results clearly show the lattice-structure dependence for
the appearance of the ferromagnetism. 
The purpose of this paper is to clarify whether
the itinerant ferromagnetism appears 
due to orbital degeneracy in infinite dimensions. 

We study the Hubbard model with doubly degenerate orbitals at each site on the 
hypercubic lattice. 
The Hamiltonian is given by
\begin{eqnarray}
  \label{2-band}
  H&=&\sum_{m=1,2}\Biggl\{
    -\dfrac{t}{2\sqrt{d}} \sum_{\langle i,j \rangle\in {\rm NN} \atop
      \sigma=\uparrow,\downarrow}
          (c_{im\sigma}^\dagger c_{jm\sigma} + h.c.)\\
& &-\mu \sum_{i,\sigma} n_{im\sigma}
   +U\sum_{i} n_{im\uparrow} n_{im\downarrow} \Biggr\} \nonumber\\
& &+U^\prime \sum_{i,\sigma,\sigma^\prime} 
                   n_{i1\sigma} n_{i2\sigma^\prime}
   -J \sum_{i,\sigma,\sigma^\prime}
          c_{i1\sigma}^\dagger c_{i1\sigma^\prime} 
          c_{i2\sigma^\prime}^\dagger c_{i2\sigma}\nonumber\\
& &-J^\prime \sum_i
         (c_{i1\uparrow}^\dagger c_{i1\downarrow}^\dagger 
          c_{i2\uparrow} c_{i2\downarrow} + h.c.)\nonumber
\end{eqnarray}
where $c_{im\sigma}$ ($c^\dagger_{im\sigma}$) denotes the annihilation 
(creation) operator of the electron at site $i$ with orbital $m$(=1 or 2) and
spin $\sigma$. The number operator is denoted by $n_{im\sigma}$.  
All the interaction terms in Eq.~(\ref{2-band}) originate from the 
Coulomb interaction between electrons at the same site.  We treat
every interaction exactly. 
Generally interaction parameters satisfy the relation 
$U\ge U^\prime \ge J\simeq J^\prime$. A symmetry argument leads to
the relation $U=U^\prime+2J$.\cite{Oles} 
Hopping terms are scaled with the dimensionality $d$ so that the 
one-electron density of states (DOS) behaves as
$D(\varepsilon)=\exp(-\varepsilon^2/t^2)/t\sqrt{\pi}$ in the limit
$d\rightarrow\infty$. 

The double-exchange mechanism due to strong Hund coupling $J$ 
was proposed to produce itinerant ferromagnetism for $n>1$, 
where $n$ denotes electron 
density per site.\cite{Slater,Zener,VVleck} 
At the 1/4-filling ($n=1$) the ground state is expected to be insulating
 in the strong-coupling limit.
In this case the effective interaction due to  virtual hoppings of electrons 
favor a ferromagnetic ground state with alternating orbital
order.\cite{Roth,KugelK,CyrotL-C,Inagaki} 
In one dimension, the ferromagnetism was 
found numerically near
$n=1$,\cite{GillS,KusakabeA2,KusakabeA1,KueiS,Hirsch} and shown
rigorously for general $n$ in the strong-coupling
limit.\cite{Kubo,KusakabeA1} For higher dimensions, it is not yet
clear whether ferromagnetic order appears. 
Inagaki and Kubo obtained ferromagnetic phases on the simple cubic lattice
for $J^\prime=0$    
using the Hartree-Fock approximation.\cite{InagakiK}
However, it is not quite clear how  
 the predicted metallic 
ferromagnetism in three dimensions is stable, 
since the the local quantum fluctuations and 
the pair-hopping ($J^\prime$) term, both of which suppress the
magnetic order, are neglected in their study. Effects of orbital
degeneracy were also studied with a $d=\infty$ two-band model in the
anisotropic (Ising) limit of 
Hund coupling and $J^\prime =0$.\cite{Rozenberg,KajueterK,HeldV} 

In this paper we study the model (\ref{2-band}) in 
infinite dimensions  near the quarter filling. We present the 
one-site effective action of the model (\ref{2-band}) and study 
the ground-state properties using the exact diagonalization method. 
We found a metallic ferromagnetic phase for electron doped cases,
e.g. $n=1.2$, which indicates that the double-exchange mechanism can
stabilize ferromagnetism in infinite dimensions. 
At the $1/4$-filling, the ground state shows a metal-insulator
transition at a finite parameter $U$. A phase diagram with both
ferromagnetic and antiferromagnetic phases is obtained for $n=1$. 
The phase diagram is consistent with a mean-field theory based on the 
perturbational treatment from the strong-coupling limit.

To derive the one-site effective action of the present model, we
employ the cavity method.\cite{GeorgesKKR} 
Taking the trace for all degrees of freedom except for those on site $0$, 
we obtain
the effective action, 
\begin{equation}
  \exp[-S_{\rm eff}(c_{0m\sigma}^\dagger,c_{0m\sigma})]
  \equiv  
  \frac{\displaystyle Z_{\rm eff}}{Z} \int \prod_{i\ne 0 \atop m,\sigma} 
  Dc_{im\sigma}^\dagger Dc_{im\sigma} e^{-S},
\end{equation}
where the partition function of the model is written as a functional
integral over Grassman variables. Generally, $S_{\rm eff}$ is an
infinite series of multiples of Grassman variables at
various times. In the $d=\infty$ limit, the action becomes
quite simple as  
\begin{eqnarray}
  S_{\rm eff} &=& - \beta \sum_{m=1,2 \atop \sigma=\uparrow,\downarrow} 
\sum_{\omega_n} 
c_{0m\sigma}^\dagger (i\omega_n) {{\cal G}_{0m\sigma}}^{-1}(i\omega_n) 
c_{0m\sigma} (i\omega_n)\nonumber\\
&+& \int^\beta_0 d\tau 
H_{\rm int} (c_{0m\sigma}^\dagger (\tau), c_{0m\sigma} (\tau))
\label{e-action}, 
\end{eqnarray}
where $H_{\rm int}$ denotes the local interaction on the site 0, and 
${\cal G}_{0m\sigma}$ satisfies the relation 
\begin{equation}
\label{sc-condition}
{\cal G}_{0m\sigma} (i\omega_n) 
= [ G_{m\sigma} (i\omega_n)^{-1} + \Sigma_{m\sigma} (i\omega_n) ]^{-1}.
\end{equation}
To derive Eqs.~(\ref{e-action}) and  (\ref{sc-condition}), we have
used the facts that the self-energy is independent of momentum and
that $\langle c_i^\dagger c_j \rangle \sim (1/\sqrt{d})^{|i-j|}$,
which applies for large $d$. 

Here we briefly explain the method for calculating the ground state. 
This one-site effective action can be approximated by the following
two-channel impurity model:
\begin{eqnarray}
  \label{impurity}
  H&=& \sum_{m=1,2\atop \sigma=\uparrow,\downarrow} \Biggl\{
       \sum_{p=2,n_s} 
             \varepsilon_{pm\sigma} a_{pm\sigma}^\dagger a_{pm\sigma} 
-\mu c_{m\sigma}^\dagger c_{m\sigma} \\
& &   +\sum_{p=2,n_s} V_{pm\sigma} 
     (a_{pm\sigma}^\dagger c_{m\sigma} + h.c.)
\Biggr\}
\nonumber\\
& & +U\sum_{m} n_{m\uparrow} n_{m\downarrow}
+U^\prime \sum_{\sigma,\sigma^\prime} n_{1\sigma} n_{2\sigma^\prime}
\nonumber\\
& &-J \sum_{\sigma,\sigma^\prime}
      c_{1\sigma}^\dagger c_{1\sigma^\prime} 
      c_{2\sigma^\prime}^\dagger c_{2\sigma}
-J^\prime (c_{1\uparrow}^\dagger c_{1\downarrow}^\dagger 
           c_{2\uparrow} c_{2\downarrow} + h.c.)\nonumber,
\end{eqnarray}
where $a_{pm\sigma}$ ($c_{m\sigma}$) denotes the annihilation operator of the
conduction (impurity) electron with channel $m$ and spin $\sigma$. 
The effective action of the impurity site has the same form as 
Eq.~(\ref{e-action}) with 
\begin{equation}
  {\cal G}_{m\sigma}^{\rm imp} (i\omega_n) = \left[ i\omega_n + \mu 
- \sum_{p=2,n_s} \frac{{V_{pm\sigma}}^2}
                      {i\omega_n - \varepsilon_{pm\sigma}} \right]^{-1}
\end{equation}
instead of ${\cal G}_{0m\sigma} (i\omega_n)$. 
Conduction electrons have $n_s-1$ orbitals. 
The effective action (\ref{e-action}) can be
well approximated with the impurity model (\ref{impurity}), 
if we increase $n_s$ and select the most suitable parameters
$(\varepsilon_{pm\sigma},V_{pm\sigma})$ so that 
${\cal G}_{m\sigma}^{\rm imp}$ satisfies the self-consistency
 condition (\ref{sc-condition}). 
We numerically diagonalize the impurity model (\ref{impurity}) with 
the Lanczos method and obtain the ground state under the
self-consistency condition (\ref{sc-condition}). It has been shown
that this method works well and the ground state can be accurately
described even with small number $n_s=5$ or $6$.\cite{GeorgesKKR}
In spatially uniform ground states 
the Green function is related to the self-energy with
$G_{m\sigma} ({\bf k}, i\omega_n)^{-1}=(G_0)^{-1}-\Sigma
-\varepsilon_k$, where $G_{0}=1/(i\omega_n+\mu)$ and  
$\varepsilon_k=d^{-1/2}\sum_{\mu=1}^{d}\cos k_\mu $. 
We also consider states with two-sublattice structures. 
The Green function from $l$ sublattice to $l^\prime$ sublattice,
$G^{ll^\prime}_{m\sigma} ({\bf k}, i\omega_n)$, has the form
\begin{equation}
\left(
\begin{array}{cc}
(G_0)^{-1}-\Sigma^A & -\varepsilon_k \\
-\varepsilon_k & (G_0)^{-1}-\Sigma^B
\end{array}
\right)^{-1}.
\end{equation} 
We have two
effective actions for two sublattices. The self-consistency condition
is unchanged as 
${\cal G}_{0m\sigma}^l (i\omega_n) =[G_{m\sigma}^{ll}(i\omega_n)^{-1}
+ \Sigma_m^l (i\omega_n)]^{-1}$,
where ${\cal G}^{l}_{0m\sigma} (i\omega_n)$ denotes the dynamical mean
field applied to the sublattice $l$. 

We study the ground states at the fillings $n=1, 1.2$, and 0.8,
controlling the chemical potential. Parameters are set as
$U=U^\prime+2J$ and $J=J^\prime$. Both spatially uniform states
and those with the two-sublattice structure are considered. 
Numerical calculations were done for $n_s = 5$ or $n_s = 6$. 
We mostly studied the system with $n_s=5$ and confirmed 
phase boundaries using the system with $n_s=6$. We found that $n_s$
dependence of the ground state is small and the phase diagrams are
the same between the cases $n_s=5$ and 6. 
\begin{figure}[htbp]
  \begin{center}
    \leavevmode
    \epsfig{file=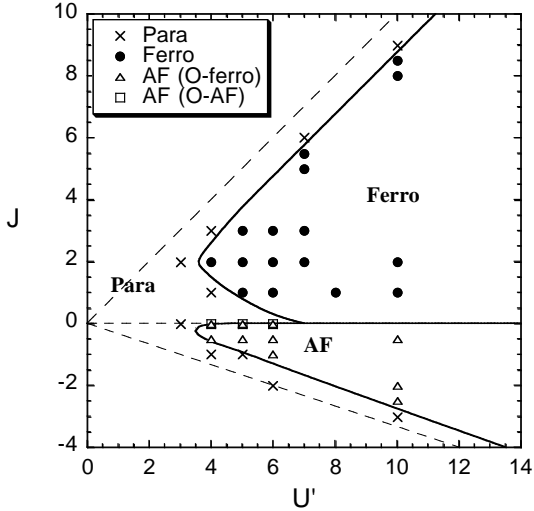,width=2.9in}
  \end{center}
    \caption{Magnetic phase diagram at the 1/4-filling ($n=1$). 
      Parameters are fixed as $U=U^\prime+2J$ and $J=J^\prime$.}
\label{fig:PD1}
\end{figure}

For the 1/4-filling, 
we found three ground-state phases, i.e., paramagnetic, ferromagnetic, 
and antiferromagnetic
ones. We obtained the phase diagram shown in Fig.~\ref{fig:PD1}. 
Though $J$ will be positive in real systems, we consider both positive
and negative $J$ to clarify the effects of Hund coupling. 
Near the phase boundary, two (or three) solutions coexist and
their energies cross over. We selected the ground state
comparing the energies and determined the phase diagram. 

\begin{figure}[bhtp]
  \begin{center}
    \leavevmode
    \epsfig{file=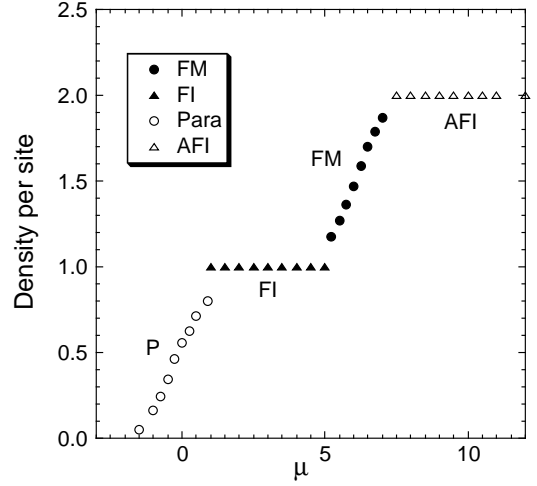,width=3.0in}
  \end{center}
    \caption{Chemical potential dependence of the electron density at
      the parameters $U=18$, $U^\prime=10$ and $J=J^\prime=4$. 
      Magnetic order
      of each state is written by FM, P, FI and AFI. FM and P denote
      ferromagnetic and paramagnetic metal, respectively. FI means
      ferromagnetic insulator with alternating orbital order and AFI
      denotes antiferromagnetic insulator with no orbital order.}
    \label{fig:density}
\end{figure}

For a wide parameter region, we found a ferromagnetic phase. This
state has an orbital super-lattice structure. 
The pseudospin $\tau_i^z=\sum_\sigma (n_{i1 \sigma}-n_{i2
  \sigma})/2$, which represents the orbital degree of freedom, 
is antiferromagnetically ordered. In this phase  
 the compressibility, $dn/d\mu$, is always vanishing and hence the 
ground state is insulating (See Fig.~\ref{fig:density}). 
Thus magnetic transition and metal-insulator transition occur at the
same coupling parameter. 
The magnetization is hardly reduced from the full polarization. 
For example, the magnetization per site($=N_e$) is 
$\langle M \rangle/N_e = 0.5000$ 
for $U=18$, $U^\prime=10$ and $J=4$, where $N_e$ denotes the number of 
electrons. Rigorously speaking the spins
cannot be  fully polarized in the $d\rightarrow\infty$ limit because
of the Gaussian DOS.\cite{FazekasMM-H} Since the reduction is too
small, it is invisible within accuracy of the present calculation. 

Another insulating phase appears for negative $J$. In this phase the
ground state has antiferromagnetic long-range order and 
also uniform orbital order.
The sublattice magnetization per site is very close to 1/2. 

The phase boundaries of the above two phases are well understood
in terms of the effective Hamiltonian in the strong-coupling limit.
We start from the ground states in the atomic limit ($t=0$) and 
 treat the hopping term as a perturbation.\cite{KugelK} 
Then the second-order perturbation leads to  
the following effective Hamiltonian for spin operators {\boldmath S}$_i$ and
pseudospin operators {\boldmath $\tau$}$_i$\cite{KugelK}: 
\begin{eqnarray}
  \label{eq:H_eff}
&&H_{\rm eff}= -\dfrac{t^2}{2d} \sum_{\langle i,j \rangle}
  \Biggl[ \dfrac{4U}{U^2-{J^\prime}^2}
  \biggl(\dfrac{1}{4}+\tau_i^z\tau_j^z\biggr) 
  \biggl(\dfrac{1}{4}-{\bf S}_i \cdot {\bf S}_j \biggr) \\
&&-\dfrac{2J^\prime}{U^2-{J^\prime}^2}
  (\tau_i^-\tau_j^- + \tau_i^+\tau_j^+)
   \biggl(\dfrac{1}{4}-{\bf S}_i \cdot {\bf S}_j\biggr) \nonumber\\
&&+\dfrac{2U^\prime}{ U^{\prime 2}-J^2 }
  \biggl\{ \dfrac{1}{4} 
   - \tau_i^z \tau_j^z 
   - 2(\mbox{\boldmath $\tau$}_i \cdot \mbox{\boldmath $\tau$}_j 
    - \tau_i^z \tau_j^z)
    (\dfrac{1}{4}+{\bf S}_i \cdot {\bf S}_j) \biggr\}\nonumber\\
&&+\dfrac{2J}{ U^{\prime 2}-J^2 }
\biggl\{ \tau_i^z \tau_j^z 
- \mbox{\boldmath $\tau$}_i \cdot \mbox{\boldmath $\tau$}_j
+ 2(\dfrac{1}{4}-\tau_i^z \tau_j^z)(\dfrac{1}{4}+ {\bf S}_i \cdot {\bf S}_j)
\biggr\} \Biggr].\nonumber
\end{eqnarray}
Clearly, spin interactions have $SO(3)$ symmetry and, 
under the conditions $U=U^\prime + 2 J$ and $J=J^\prime$, pseudospin
interactions also 
have $SO(2)$ rotational symmetry about the $y$-axis. The derivation of
Eq.~(\ref{eq:H_eff}) is valid for $-U^\prime/3 < J < U^\prime$, where 
intermediate states have higher energy than the unperturbed ground 
states. 
The perfect ferromagnetic state with perfect 
antiferromagnetic pseudospin order has the energy  
$-t^2 /(2U^\prime - 2J)$ per site, while that of
the N\'eel state with perfect ferromagnetic pseudospin order is
$ - t^2 U /(2U^2 - 2J^2)$. Thus the ferromagnetic state has a lower
energy for $0<J<U^\prime$ and the antiferromagnetic state is lower for 
$-U^\prime/3<J<0$. 
We see in Fig.~\ref{fig:PD1} that the phase boundaries are close
to the three lines,
$J=-U^\prime/3$, $J=0$ and $J=U^\prime$.
The present result reflects the fact that the mean-field theory is
exact in infinite-dimensional localized spin models. 

\begin{figure}[thbp]
  \begin{center}
    \leavevmode
    \epsfig{file=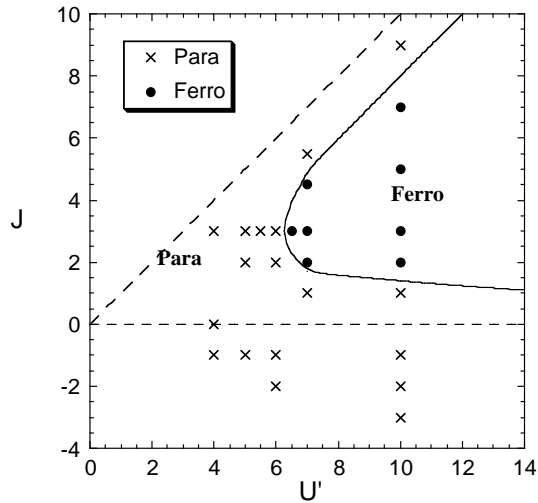,width=2.9in}
  \end{center}
    \caption{Magnetic phase diagram at the filling $n=1.2$. 
      Parameters are fixed as $U=U^\prime+2J$ and $J=J^\prime$.}
\label{fig:PD2}
\end{figure}

In order to examine the effects of electron doping to the ordered states
at the $1/4$-filling, we studied the case with  $n=1.2$. 
At this filling we obtained two metallic phases, both of which are
spatially uniform, i.e.\ the ferromagnetic phase and the
paramagnetic one (see Fig.~\ref{fig:PD2}). 
We could not find any antiferromagnetic phase at this
filling. The transition from the paramagnetic phase to the
ferromagnetic one is of first order. Both ferromagnetic and
paramagnetic states exist as solutions of the 
self-consistency equations in a region close to
the phase boundary and their energies cross over at the phase boundary. 

The ferromagnetic state is metallic, since it has  finite compressibility
(Fig.~\ref{fig:density}), and it has no orbital ordering. 
The reduction of magnetization is
larger than the insulating ferromagnet at the $1/4$-filling. For
example, $\langle M \rangle/N_e=0.341$ for $U=18$, $U^\prime=10$, and
$J=4$ at $n=1.2$. 
The area of the ferromagnetic phase is reduced in the phase diagram
compared to that of the insulating
ferromagnetic phase at the 1/4-filling.
On the other hand, the
antiferromagnetic phase completely disappears at this filling.
This result is consistent with the well-known fact that the itinerancy of 
electrons (or holes) is  hardly compatible   
with the antiferromagnetic long-range order. 

As a hole-doped case, we study the ground state for
$n=0.8$. At this filling we found only metallic ground states which are 
translationally invariant. We could not
find any magnetically ordered phase for $J\le U^\prime \le 20$.
Disappearance of antiferromagnetism is understood 
as the result of the itinerancy of the holes.  
Self-consistent solutions that show ferromagnetism also appear for
large $U^\prime$, e.g. $U^\prime=10$, but have slightly higher energy
than that of the paramagnetic states. 
Since the probability of double occupancy is low  
for $n<1$, the double-exchange mechanism due to Hund coupling is  
less effective than in the case with $n>1$ and hence ferromagnetism 
may disappear.
The ferromagnetic ground state however appears even for $n<1$ 
in one dimension.\cite{KusakabeA1,Hirsch} 
This discrepancy may be understood with the argument that higher 
dimensionality destroys the perfect
ferromagnetism,\cite{FazekasMM-H,Edwards} though we need more 
study to conclude the absence of the metallic ferromagnetism 
at less than $1/4$-filling for strong coupling. 

In this study, we found the metallic ferromagnetism, in which spins
are partially polarized, for positive $J$
only at more-than $1/4$-filling ($n=1.2$). This result should be
compared with the single-band Hubbard model, which does not show
ferromagnetism.\cite{FreericksJ} This indicates that the 
double-exchange mechanism due to the Hund-coupling
favors ferromagnetism  and realizes metallic
ferromagnetism for strong but finite coupling in infinite dimensions. 
The ferromagnetism appears in a parameter region where $0<J<U^\prime$.
We did not search for ground states with larger sublattice structures
in the present study. 
Hence possibility of incommensurate states still remains. 
Further extension of the study in infinite dimensions and 
also a reliable phase diagram of  
the metallic ferromagnetism in one dimension, which is still not available, 
will be useful to gain insight into the role of the orbital degeneracy 
in the metallic ferromagnetism in two and/or three dimensions. 

We would like to thank Dai Hirashima, Koichi Kusakabe, Tetsuya Mutou,
and Harumi Sakamoto for stimulating discussions and comments. 
This work was supported by Grant-in-Aid No.\ 09640453 from the Ministry of 
Education, Science and Culture of Japan. 
The numerical calculations were done on 
Facom VPP500 at the ISSP of the University of Tokyo and DEC Alpha 500
personal computer at the Institute of Physics of the University of
Tsukuba.

\end{document}